\documentclass{article}
\usepackage{spconf,amsmath,graphicx}
\usepackage{multirow}
\usepackage{graphics}
\usepackage{balance}
\usepackage{url}

\title{Perceptual Loss Function for Neural Modelling of Audio Systems}
\name{Alec Wright and
    Vesa V\"alim\"aki\sthanks{This research is part of the activities of the Nordic Sound and Music Computing Network---NordicSMC (NordForsk project no.~86892).}}
\address{Acoustics Lab, Dept. Signal Processing and Acoustics, Aalto University, FI-02150 Espoo, Finland\\ alec.wright@aalto.fi\\ }
\begin{document}
%
\maketitle
\begin{abstract}
This work investigates 
alternate pre-emphasis filters used as part of the loss function during neural network training for nonlinear audio processing. In our previous work, the error-to-signal ratio loss function was used during network training, with a first-order highpass pre-emphasis filter applied to both the target signal and neural network output. This work considers more perceptually relevant pre-emphasis filters, which include lowpass filtering at high frequencies. We conducted listening tests to determine whether they offer an improvement to the quality of a neural network model of a guitar tube amplifier. Listening test results indicate that the use of an A-weighting pre-emphasis
filter offers the best improvement among the tested filters. The proposed perceptual loss function improves the sound quality of neural network models in audio processing without affecting the computational cost.
\end{abstract}
\section{Introduction}
\label{sec:intro}

Virtual analog modelling is a field of research which seeks to create algorithms that emulate music hardware, such as instruments, amplifiers or audio effects \cite{Valimaki11}. Music hardware, specifically guitar amplifiers and distortion effects, exhibit highly nonlinear behaviour, which is particularly difficult to emulate accurately \cite{pakarinen2009review}. This study is on the topic of virtual analog modelling for highly nonlinear audio circuits, using Recurrent Neural Networks (RNNs).

Approaches to virtual analog modelling exist on a scale between ``white-box'' and ``black-box''. In ``white-box'' modelling the physical characteristics of the circuit or instrument are studied and used to derive equations that describe their behaviour \cite{Karjalainen06, santagata2007non, Paiva12, werner:2015}. In ``black-box'' modelling the input-output relationship of the device is emulated directly, by using data measured from the device \cite{novak2010chebyshev, orcioni2018identification}. ``Grey-box'' models fall somewhere between these two approaches \cite{kiiski2016time,eichas2018gray}. 

In recent years a number of researchers have published work on black-box modelling of nonlinear audio circuits using neural networks. This has included a feedforward WaveNet-style model \cite{damskaggICASSP,damskaggSMC}, as well as RNNs \cite{zhang2018vacuum, wright:2019} and a hybrid model, which combined a convolutional layer with an RNN \cite{schmitz2018nonlinear}. In our previous work \cite{damskaggICASSP, damskaggSMC, wright:2019} the \textit{error-to-signal ratio} (ESR) loss function was used during network training, with a first-order highpass \textit{pre-emphasis} filter being used to suppress the low frequency content of both the target signal and neural network output. 

In this work we investigate the use of alternate pre-emphasis filters in the training of a neural network consisting of a Long Short-Term Memory (LSTM) and a fully connected layer. We have conducted listening tests to determine whether the novel loss functions offer an improvement to the quality of the resulting model. This paper shows that in addition to the highpass filtering at low frequencies, it is advantageous to suppress the highest audio frequencies in the loss function, since such high frequencies are perceptually irrelevant. 

The rest of this paper is structured as follows. Sec.~2 introduces the model used during this work. Sec.~3 introduces the device being modelled and the training procedure.
Sec.~4 describes the pre-emphasis filters used during this work. Sec.~5 compares the resulting models, both in terms of objective error measurements and in terms of listening test scores. Finally, Sec.~6 concludes the paper.

\section{Neural Network Model}
\label{sec:model}

\begin{figure}[t!]
    \includegraphics[width=0.93\columnwidth]{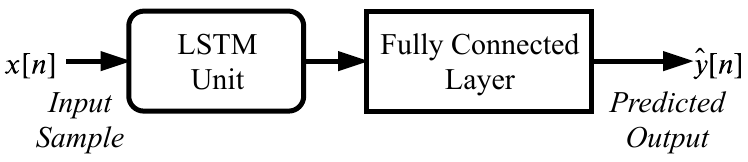}
    \caption{{\it Neural network architecture, where $x[n]$ is the input signal and $\hat{y}[n]$ is the network's predicted output sample.}}
    \label{fig:model}
\end{figure}

 The model used in this work was first proposed in \cite{wright:2019}, it is depicted in Fig.~{\ref{fig:model}}. The model consists of a single LSTM unit, followed by a fully connected layer. It is an end-to-end model which takes a sequence of audio samples as input, and produces a sequence of audio samples as output. At each discrete time step in the sequence, the model predicts the corresponding output sample produced by the modelled device, based on the input sample, LSTM state and the model's parameters:
\begin{equation}
	\hat{y}[n] = f(x[n], s[n-1], \theta),
	\label{eq:model}
\end{equation}

\noindent where $n$ is the discrete time index, $\hat{y}[n]$ and $x[n]$ are the RNN's predicted output sample and the corresponding unprocessed input sample respectively, $s[n-1]$ is the LSTM state from the previous time step, and $\theta$ are the RNN's parameters which are learned during training.

\section{Modelled Device and Training}
\label{sec:training}
	
In this study, the RNN models were trained to emulate the Blackstar HT-1 guitar amplifier \cite{blackstar}, which was modelled previously in \cite{wright:2019}. This is a small 1-Watt vacuum tube amplifier, with both a high gain and a low gain channel. The models were trained using guitar and bass audio processed by the amplifier. The dataset consisted of the same material used in \cite{wright:2019}, a total of approximately seven minutes of audio. The dataset was recorded with the amplifier set to the high gain channel, with the ``EQ'' and ``volume'' controls set to 10 and the ``gain'' control set to 5. All audio used in this study was recorded at a sampling rate of 44.1 kHz. 

The training process was similar to that described in \cite{wright:2019}, with the data being split into half second segments and processed in mini-batches. For each segment, the first 1000 samples would be processed without gradient tracking, to allow the LSTM state to initialise. Then the remainder of the segment could be processed, with backpropagation and parameter updates being carried out every 2048 samples. After the complete training dataset was processed, it would be shuffled before the next training epoch began. All models were trained for a total of 750 epochs.

One issue with the comparison of pre-emphasis filters is that, due to random nature of neural network training, two networks of identical structure, once trained, will achieve different errors on the test set. When comparing two networks trained using different pre-emphasis filters, it is not clear if any differences in performance are due to the pre-emphasis filter used, or occur as a result of the training process. To compensate for this, five copies of each RNN configuration were trained, with the RNN that achieved the lowest loss, i.e.~the best result, on the test set being used for evaluation.
	
\section{Loss Function Pre-Emphasis Filtering}
\label{sec:lossfunc}

The neural networks were trained to minimise a loss function representing the ESR, which is the squared error divided by the energy of the target signal. The aim of this study was to investigate the application of different pre-emphasis filters to the network output and target signals, prior to calculating the loss. The aim of applying a pre-emphasis filter is to emphasise certain frequencies in the loss function, thus teaching the network to prioritise minimising certain frequencies over others. For an audio sequence of length $N$, the pre-emphasised ESR loss is given by
\begin{equation}
	\mathcal{E_{\text{ESR}}} = \frac{\sum_{n=0}^{N-1} |y_p[n] - \hat{y}_p[n]|^2}{\sum_{n=0}^{N-1}|y_p[n]|^2}  ,
	\label{loss}
\end{equation}

\noindent where $y_p$ is the pre-emphasised target signal and $\hat{y}_p[n]$ is the pre-emphasised output of the neural network. As proposed in \cite{wright:2019}, a loss function measuring the difference in DC offset between the target and network output signals was also included, and is given by
\begin{equation}
    \mathcal{E}_{\text{DC}} = \frac{|\frac{1}{N}\sum_{n=0}^{N-1}(y[n]-\hat{y}[n])|^2}{\frac{1}{N}\sum_{n=0}^{N-1}|y[n]|^2}.
\end{equation}

\noindent The loss function used during training is simply the sum of the two loss terms:
\begin{equation}
    \mathcal{E} = \mathcal{E_{\text{ESR}}} + \mathcal{E}_{\text{DC}}.
\end{equation}

\begin{figure}[t!]
    \includegraphics[width=\columnwidth]{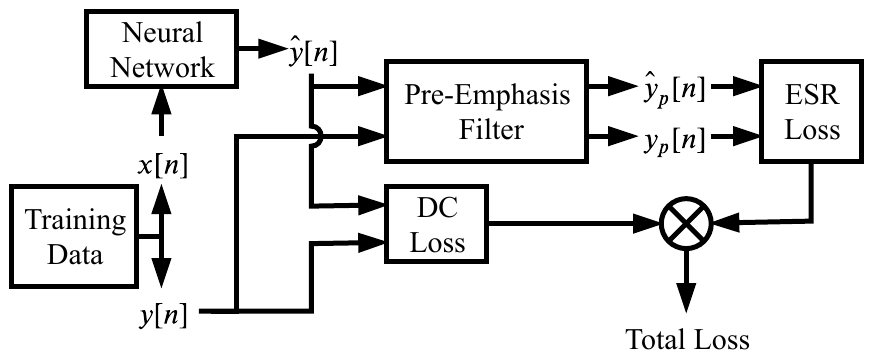}
    \caption{{\it Training process including the loss calculations.}}
    \label{fig:train}
\end{figure}

\noindent The process of calculating the loss is depicted in Fig.~{\ref{fig:train}}. The objective of this study is to investigate how the perceptual quality of the model output is affected by the choice of loss function pre-emphasis filter. The rest of this section describes the filters tested.

\subsection{First-order highpass}
\label{ssec:FOHP}

As proposed in \cite{damskaggICASSP}, and used in \cite{damskaggSMC,wright:2019}, a first-order highpass pre-emphasis filter was the first pre-emphasis filter we considered during this study. Its  transfer function is
\begin{equation}
    H_{HP}(z) = 1 - 0.85 z^{-1}.
\end{equation}

\noindent The magnitude response of this filter is shown in the top panel of Fig.~\ref{fig:FR}. This type of filter is commonly used in speech processing to boost spectral flatness \cite{markellinearpred}. In previous work this has been intended to suppress low frequencies in the signal, as the energy in the guitar amplifier output tends to be more concentrated in the lower end of the frequency spectrum.

\subsection{Folded differentiator}
\label{sssec:FoldDiff}

The second pre-emphasis filter we tested during this study is the folded differentiator filter, with the transfer function
\begin{equation}
    H_{FD}(z) = 1 - 0.85 z^{-2}.
\end{equation}

\noindent This was chosen as it was noted that the output of the guitar amplifier is very limited above 10 kHz. It is thought that attenuating these higher frequencies as well as the low frequencies, as shown in Fig.~\ref{fig:FR} (middle), might cause the network to concentrate more on the mid frequency range, which is perceptually significant.

\subsection{A-weighting filter}
\label{sssec:AWeight}

The final pre-emphasis filter we tested during this study is based on the A-weighting curve, as defined in the International Electrotechnical Commission standard \cite{isoAw}. The A-weighting curve is intended to compensate for the relative loudness of different frequencies as perceived by the human ear. The A-weighting filter was then followed by a first order lowpass filter with the transfer function

\begin{equation}
    H(z) = 1 + 0.85 z^{-1},
\end{equation}

\noindent to decrease the emphasis on the high frequency region where relatively little energy is present. In this paper the low-passed A-weighting filter  will be represented as $H_{AW}$. 

\begin{figure}[t!]
    \includegraphics[width=\columnwidth]{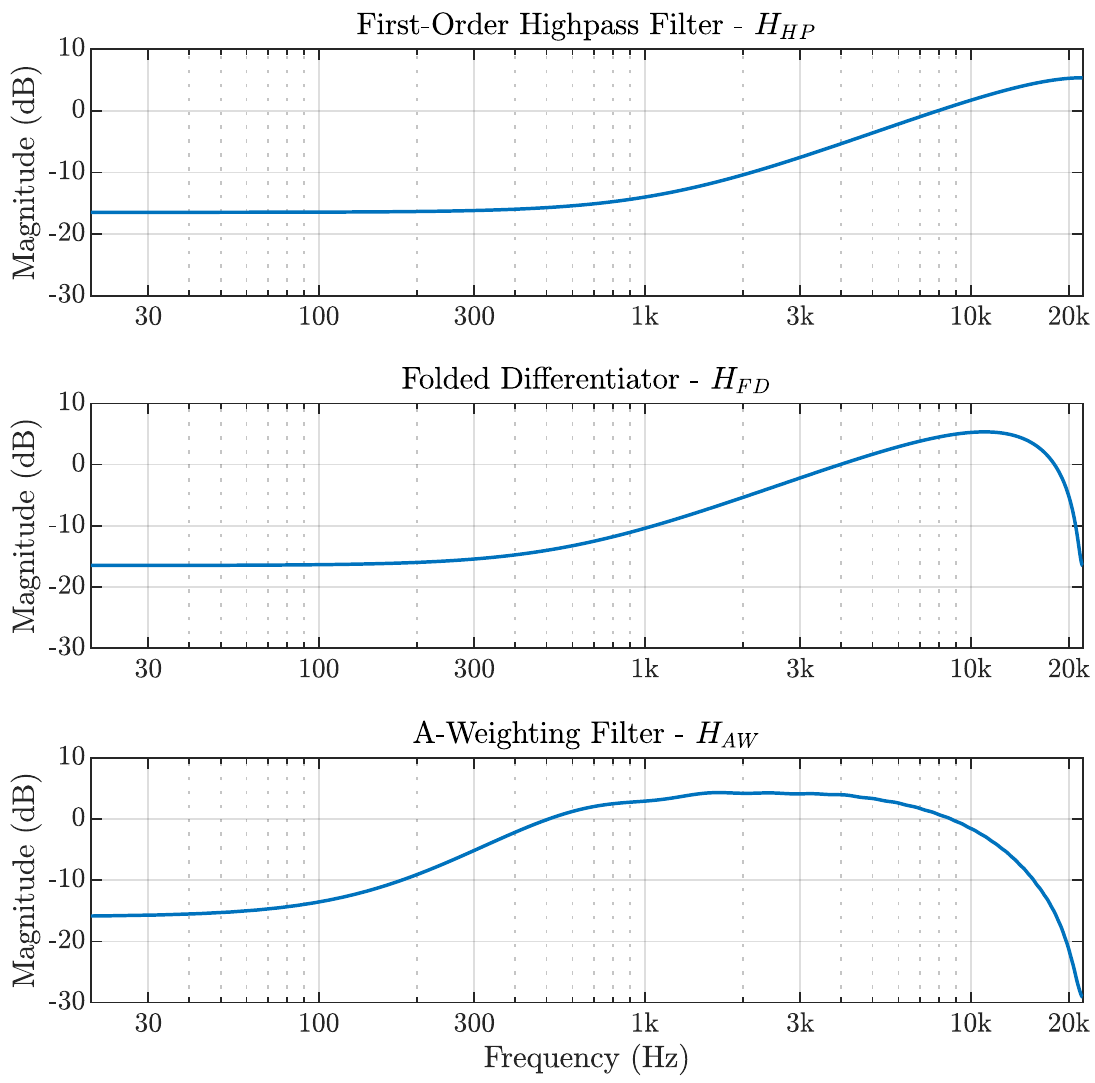}
    \caption{{\it Magnitude frequency response of pre-emphasis filters, at 44.1-kHz sample rate.}}
    \label{fig:FR}
\end{figure}

Whilst an A-weighting filter can be implemented, for example, as a eleventh-order infinite impulse response filter \cite{AwghtDesign}, for this work we chose to implement it as a 100-tap finite impulse response (FIR) filter. As FIR filters are fully parallelisable, they can be applied very cheaply on a Graphics Processing Unit (GPU), and in practice can simply be implemented as a one-dimensional convolutional layer. As such the application of an 100-tap FIR filter had a negligible effect of the overall time required to train the network. The filter was designed using the least-squares method, with the target magnitude frequency response calculated using the weighting function described in the IEC standard \cite{isoAw}. The low-passed A-weighting filter's magnitude response is shown at the bottom of Fig.~\ref{fig:FR}.
	
\section{Evaluation}
\label{sec:expdes}

To evaluate the pre-emphasis filters, the RNN model described in Sec.~\ref{sec:model} was trained with an LSTM hidden size of either 32 or 64. For both of these hidden sizes, a total of four neural networks were compared, one trained with each of the three different pre-emphasis filters described in Sec.~\ref{sec:lossfunc}, as well as one which was trained using no pre-emphasis filter. 

In our previous work we measured the running speed of a C++ implementation of this model, on an Apple iMac with a 2.8 GHz Intel Core i5 processor \cite{wright:2019}. The time required to process a second of audio for the models with LSTM hidden size of 32 and 64 was 0.12~s and 0.24~s respectively. It should be noted that as the pre-emphasis filters are only applied during training, they have no influence on the running speed of the models. 

\subsection{Objective evaluation}
\label{ssec:objeval}

For the objective evaluation, the loss was calculated for an unseen test set of electric guitar and bass guitar input signals. For each network evaluated, the test loss was calculated with the loss function with which that network was trained, as well as with the other three loss functions tested. The resulting test losses are shown in Table \ref{table:results}.

The results show that as you would expect, the RNN model with an LSTM hidden size of 64 achieved a lower test loss than the model with hidden size of 32. It might also be expected that for each loss function, the network that achieved the lowest test loss would be the network that was trained to minimise that same loss function. However, the results show that the networks trained using the $H_{FD}$ pre-emphasised loss function achieved a higher $H_{FD}$ pre-emphasised test loss, i.e.~they performed worse, than the networks trained using the $H_{HP}$ pre-emphasised loss function. Because of this, the networks trained with the $H_{FD}$ pre-emphasised loss function were not used in the listening tests.

\begin{figure}[t!]
    \includegraphics[width=\columnwidth]{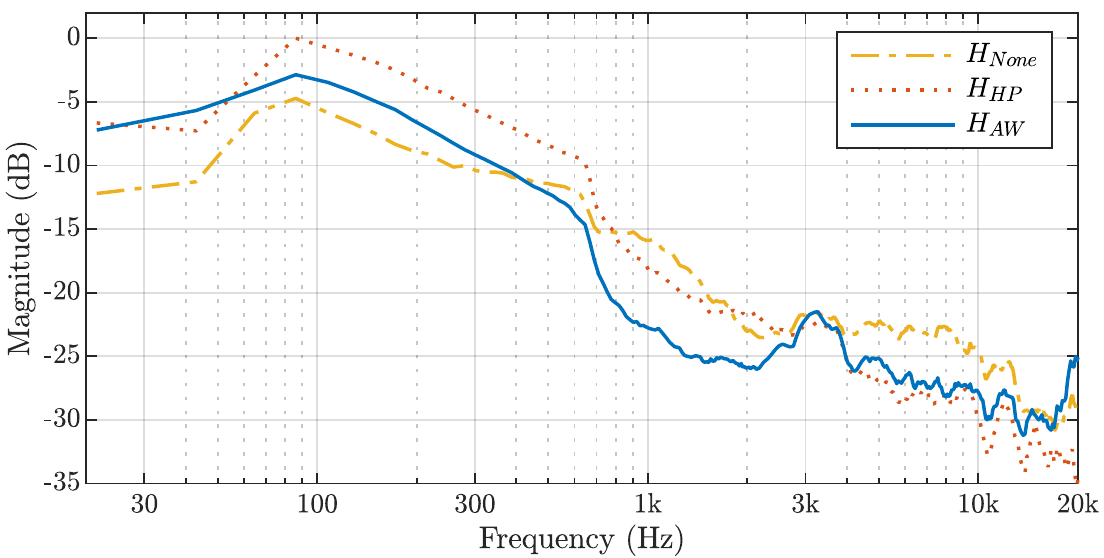}
    
    \caption{{\it Test set error in the frequency domain}}
    \label{fig:Error}
\end{figure}

The test set error signal, obtained by subtracting the network's predicted output signal from the target signal, is plotted in the frequency domain in Fig.~\ref{fig:Error}. It can be observed that the error for each network varies according to frequency. For the network trained using the $H_{AW}$ pre-emphasised loss function, the error is approximately 5 dB less than the other networks over the 1-2~kHz range, which is a region where human hearing is particularly sensitive.

\begin{table}
\centering
\caption{{\it Test loss for the RNN models}}
\vspace*{1mm}
\resizebox{\columnwidth}{!}{%
\begin{tabular}{c|c|cccc}
Hidden             & Training     & \multicolumn{4}{c}{Test Loss Pre-Emphasis}
\\ \cline{3-6}
Size                & Pre-Emph. & $H_{None}$ & $H_{HP}$ & $H_{FD}$ & $H_{AW}$                \\ 
\hline
\multirow{4}{*}{32} & $H_{None}$   & \textbf{1.56\%} & 10.9\% & 7.68\% & 3.16\%                     \\
                    & $H_{HP}$ & 2.49\% & \textbf{6.78\%} & \textbf{4.54\%} & 3.10\%                      \\
                    & $H_{FD}$ & 2.92\% & 10.1\% & 5.57\% & 3.33\%                      \\
                    & $H_{AW}$ & 3.20\% & 23.1\% & 7.73\% & \textbf{2.66\%}  \\  
\hline
\multirow{4}{*}{64} & $H_{None}$   & \textbf{0.77\%} & 5.65\% & 3.56\% & 1.41\%                      \\
                    & $H_{HP}$ & 1.08\% & \textbf{3.79\%} & \textbf{2.61\%} & 1.66\%                      \\
                    & $H_{FD}$ & 1.31\% & 5.92\% & 2.87\% & 1.63\% \\
                    & $H_{AW}$ & 1.01\% & 12.3\% & 4.15\% & \textbf{1.28\%}  
\end{tabular}
}

\label{table:results}
\end{table}

\subsection{Listening test}
\label{ssec:listeningtest}

To assess how well each of the neural networks emulate the guitar amplifier, a multiple stimuli with hidden reference and anchor (MUSHRA) listening test was conducted \cite{ITUMUSHRA}. The webMUSHRA interface was used to conduct the tests \cite{schoeffler2018webmushra}. 

For each MUSHRA trial, the participant was presented with an audio clip that had been processed through the guitar amplifier, as well as clips that had been processed by neural network models trained by each of the three loss functions being tested. Additionally an anchor, created by processing the input through a $tanh$ nonlinearity, was included. The users were asked to assign a score out of 100 to each of the test items, based on how accurately they emulated the reference.

Four different input sounds were used during the test. The sounds were clips of bass guitar, taken from the test set used during the objective evaluation. Bass guitar sounds were chosen over guitar sounds, as preliminary listening tests indicated that they resulted in differences between the RNN models and the reference that were much more audible.

Trials were conducted separately for both of the LSTM hidden sizes being tested. This resulted in a total of eight MUSHRA trials in each test. In total, 14 people, with no reported hearing problems, participated in the listening tests. The tests were conducted in sound-proof booths using Sennheiser HD-650 headphones.

\begin{figure}[t!]

    \includegraphics[width=\columnwidth]{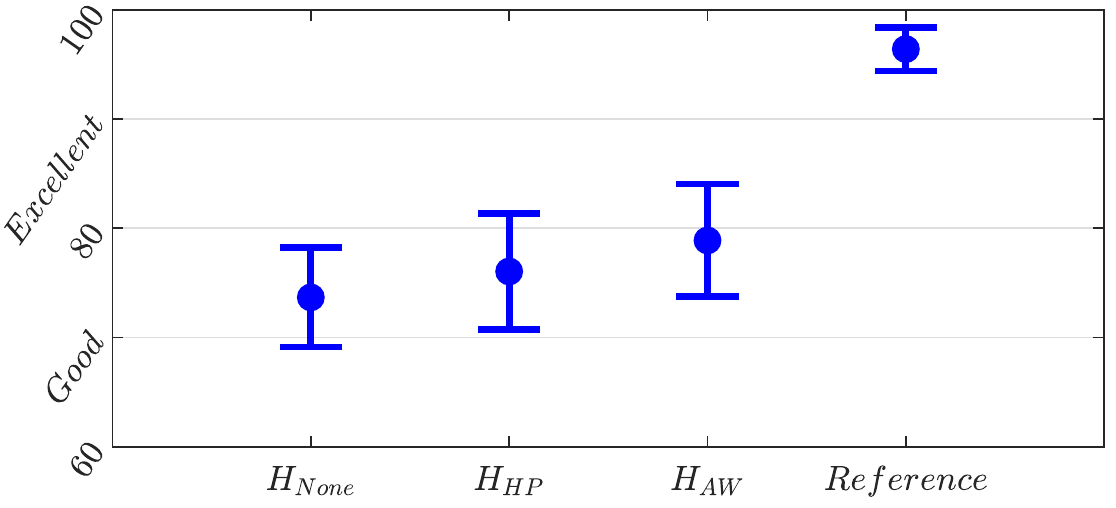}
    \includegraphics[width=\columnwidth]{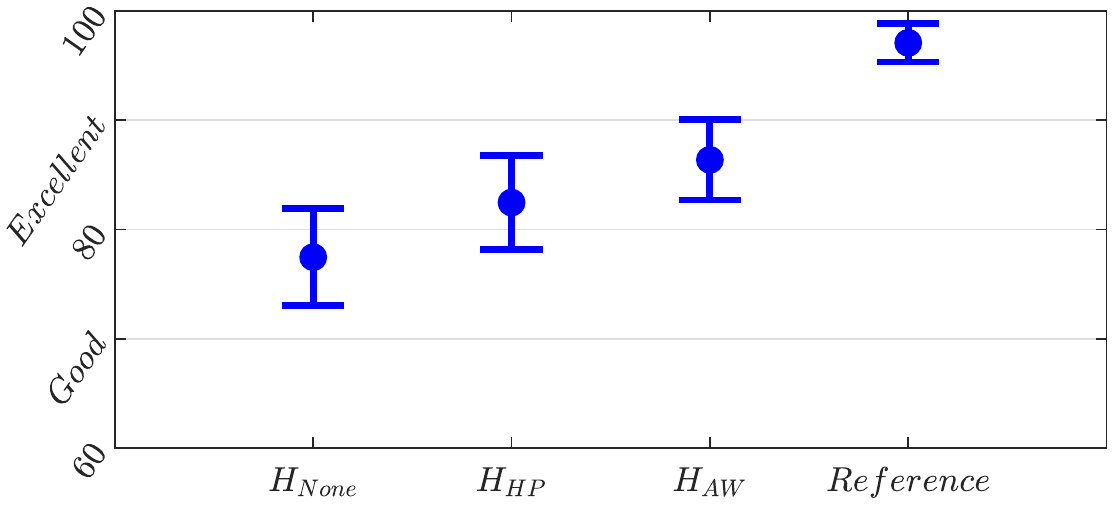}

    \caption{{\it Mean results of MUSHRA test with 95\% confidence interval, for LSTM hidden size 32 (top) and 64 (bottom).}}
    \label{fig:ListenTest2}
\end{figure}

The results of the listening test are shown in Fig.~\ref{fig:ListenTest2}, with the mean score for each method and the 95\% confidence interval being shown. The anchor achieved a mean score of 3 out of 100, but has not been included in the plot so that the test results can be observed more clearly. 
For both the LSTM hidden sizes tested, a trend can be observed, with the no pre-emphasis filtering case performing worst, and the $H_{HP}$ and $H_{AW}$ pre-emphasised cases performing second best and best respectively. 
For the models with an LSTM hidden size of 64, the use of the $H_{AW}$ pre-emphasised loss over the non pre-emphasised loss results in a statistically significant improvement in perceived emulation quality. However, in the other cases the differences are not statistically significant.

\begin{table}
\centering
\caption{{\it Mean results of MUSHRA listening test}}
\vspace*{1mm}

\resizebox{0.58\columnwidth}{!}{%
\begin{tabular}{c|ccc}
Hidden & \multicolumn{3}{c}{Mean MUSHRA Score}            \\ 
\cline{2-4}
Size   & $H_{None}$ & $H_{HP}$ & $H_{AW}$  \\ 
\hline
32     & 74              & 76            & \textbf{79}             \\
64     & 77              & 82            & \textbf{86}     
\end{tabular}
}

\label{table:MUSH}
\end{table}

\section{Conclusion}
\label{sec:majhead}

This work has evaluated the use of different pre-emphasis filters during the training of a previously proposed RNN model, used for nonlinear audio circuit modelling. The pre-emphasised loss achieved by each training condition was compared. Listening test results indicate that the use of a pre-emphasis filter can improve the model's perceptual similarity to the target device, with the A-weighting filter performing best. As the pre-emphasis filter is only used during training, the improvement in perceptual similarity comes at no extra computational cost when running the resulting model.


%
%
%


\balance
\interlinepenalty=10000
\bibliographystyle{IEEEbib}
\bibliography{strings,refs}

\end{document}